\documentclass[letterpaper, 10 pt, conference]{ieeeconf}
\IEEEoverridecommandlockouts

\title{\LARGE \bf
Attention Meets UAVs: \\ A Comprehensive Evaluation of DDoS Detection in Low-Cost UAVs
}

\author{Ashish Sharma$^{1}$, SVSLN Surya Suhas Vaddhiparthy$^{2}$,  Sai Usha Goparaju$^{3}$,\\ Deepak Gangadharan$^{4}$ and Harikumar Kandath$^{5}$
\thanks{}
\thanks{$^{1}$Ashish Sharma, Computer Systems Group, International Institute of Information Technology Hyderabad, India. {\tt\small ashish.s@research.iiit.ac.in}}%
\thanks{$^{2}$SVSLN Surya Suhas Vaddhiparthy, Computer Systems Group, International Institute of Information Technology Hyderabad, India. {\tt\small suhas.vaddhipar@research.iiit.ac.in}}%
\thanks{$^{3}$Sai Usha Goparaju, Computer Systems Group, International Institute of Information Technology Hyderabad, India. {\tt\small ushasai.g@research.iiit.ac.in}}%
\thanks{$^{4}$Deepak Gangadharan, Computer Systems Group, International Institute of Information Technology Hyderabad, India.
        {\tt\small deepak.g@iiit.ac.in}}%
\thanks{$^{5}$Harikumar Kandath, Robotics Research Center, International Institute of Information Techno
logy Hyderabad, India.
        {\tt\small harikumar.k@iiit.ac.in}}%
}

\usepackage{cite}
\usepackage{amsmath,amssymb,amsfonts}
\usepackage{hyperref}
\usepackage{algorithm}
\usepackage{algorithmic}
\usepackage{graphicx}
\usepackage{textcomp}
\usepackage{xcolor}
\usepackage{caption}
\usepackage{subcaption}
\usepackage{tabularx,booktabs}
\usepackage{pgfplots}
\pgfplotsset{compat=1.18} 
\usepackage{multirow}
\def\BibTeX{{\rm B\kern-.05em{\sc i\kern-.025em b}\kern-.08em
    T\kern-.1667em\lower.7ex\hbox{E}\kern-.125emX}}

\begin{document}

\maketitle
\thispagestyle{empty}
\pagestyle{empty}

\begin{abstract}

This paper explores the critical issue of enhancing cybersecurity measures for low-cost, Wi-Fi-based Unmanned Aerial Vehicles (UAVs) against Distributed Denial of Service (DDoS) attacks. In the current work, we have explored three variants of DDoS attacks, namely Transmission Control Protocol (TCP), Internet Control Message Protocol (ICMP), and TCP + ICMP flooding attacks, and developed a detection mechanism that runs on the companion computer of the UAV system. As a part of the detection mechanism, we have evaluated various machine learning, and deep learning algorithms, such as XGBoost, Isolation Forest, Long Short-Term Memory (LSTM), Bidirectional-LSTM (Bi-LSTM), LSTM with attention, Bi-LSTM with attention, and Time Series Transformer (TST) in terms of various classification metrics. Our evaluation reveals that algorithms with attention mechanisms outperform their counterparts in general, and TST stands out as the most efficient model with a run time of ~0.1 seconds. TST has demonstrated an F1 score of 0.999, 0.997, and 0.943 for TCP, ICMP, and TCP + ICMP flooding attacks respectively. In this work, we present the necessary steps required to build an on-board DDoS detection mechanism. Further, we also present the ablation study to identify the best TST hyperparameters for DDoS detection, and we have also underscored the advantage of adapting learnable positional embeddings in TST for DDoS detection with an improvement in F1 score from 0.94 to 0.99.

\end{abstract}


\section{Introduction}
\label{Introduction}

Technological innovations have recently spread their wings through Unmanned Aerial Vehicles (UAVs), marking a dawn of advancements in aerial ingenuity. These advances have resulted in unparalleled growth of UAV adaptation, particularly low-cost UAVs, which are anticipated to reach USD 42.8 billion in the market by 2025~\cite{drones5020046}. The autonomy and cost-effectiveness combined with the agile nature of low-cost UAVs have drastically increased their visibility in various applications, such as Traffic Management, Healthcare, Disaster Response and  Mega Sporting Events~\cite{al2023systematic}.


As the use of UAVs proliferates across various applications, scalability and reliable connectivity requirements become paramount for a UAV-based approach. Amidst the available approaches, Wi-Fi-based UAV connectivity emerges as one potential solution to enable the full capabilities of a UAV system~\cite{7805730}. One can establish resilient networks supporting swarm deployments by leveraging Wi-Fi technology~\cite{shrit2017new}.

\begin{figure}[hbt]
  \centering
  \includegraphics[width=0.90\columnwidth]{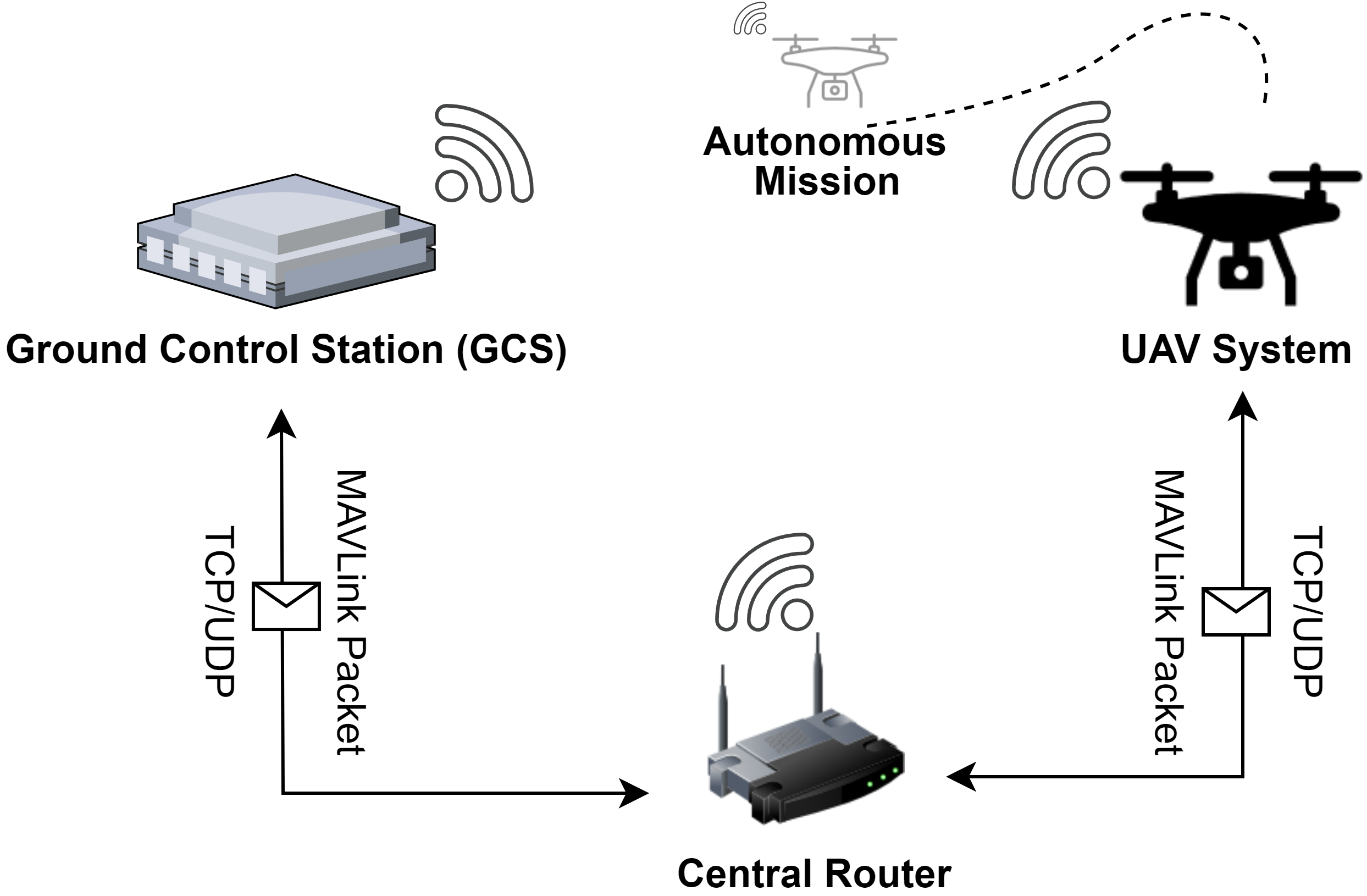}
  \caption{General Wi-Fi-based UAV System}
  \label{Figure1}
\end{figure}

Figure \ref{Figure1} illustrates a general Wi-Fi-based UAV system involving a Central Router, Ground Control Station (GCS), and a UAV system. A typical UAV system uses a Micro-Air- Vehicle link (MAVlink), a lightweight open-source communication protocol for seamless two-way communication between the GCS and the UAV system. MAVlink adapts a binary serialization approach for overhead-free communication between the devices in the network. Despite its widespread usage and development, this protocol is vulnerable to numerous security attacks, such as  
spoofing, Denial of Service (DoS), and message foraging attacks \cite{sabuwala2023approach} \cite{8743355}.

According to Esentire's 2023 Official Cybercrime report \cite{esentire}, the global annual cost of cybercrime is predicted to reach USD 9.5 trillion in 2024  and is projected to reach USD 10.5 trillion by 2025. The report highlights that this heightened risk can pertain to the explosion in mobile, cloud, Internet of Things (IoT) usage, and remote tools. The Cybersecurity statistics report by Cobalt \cite{cobalt} highlights that Distributed Denial of Service (DDoS), Ransomware, and Phishing attacks hold a predominant share as the first-hand choices of the attackers. The report particularly emphasizes that resource constraint edge applications such as IoT devices and UAVs fall prey to DDoS attacks significantly, making DDoS one of the most calamitous forms of cyber security attack \cite{10115435}.  

On the other hand, multiple stakeholders opted to integrate AI with cybersecurity detection \cite{8743355} in the endeavor to find a solution. Nevertheless, these efforts were promptly halted as this approach raised privacy and security issues according to the General Data Protection Regulation (GDPR) \cite{cobalt}, as the detection models required data to be transferred outside the resource-constrained system for processing. Considering the UAV system's resource-constrained nature, it is vital to deploy a resource-efficient mechanism on the UAV system to hinder any potential DDoS attacks. A general-purpose UAV system is equipped with a companion computer \cite{8743355} capable of running a routine or a lightweight task. Therefore, an efficient approach can involve deploying a DDoS Detection model on the companion computer of the UAV system. Based on this notion, in the current work, we conduct a comprehensive evaluation considering various algorithms for DDoS Detection.  

\textit{\textbf{The contributions of our work are as follows.}}

\begin{enumerate}

    \item We present the necessary pre-processing steps to collect a comprehensive dataset while conducting flooding attacks on UAVs, as there is a scarcity of diverse datasets tailored explicitly for such attacks. 

    \item We evaluate the feasibility of deploying Machine
Learning (ML) and Deep Learning (DL) architectures
on resource-constrained UAV companion computers
for DDoS detection. Specifically, we have evaluated
various algorithms, such as XGBoost (XGB), Isola-
tion Forest (IF), Long Short-Term Memory (LSTM),
Bidirectional-LSTM (Bi-LSTM), LSTM with attention
(LSTM-A), Bi-LSTM with attention (BLSTM-A), and
Time Series Transformer (TST) in terms of various
classification metrics against three flooding attacks:
TCP, ICMP, and TCP with ICMP.

    \item To the best of our knowledge, for the first time, we have explored the capability of TST and compared it with other attention mechanisms for DDoS attack detection in UAVs. We have also conducted an ablation study to identify the best TST hyperparameters and presented the significance of learnable-positional embeddings in improving the F1 score.
    



    

\end{enumerate}

\section{Related Works}
\label{Related Works}

This section presents the UAV network attack scenario, followed by literature on DDoS attacks and detection mechanisms.

\subsection{UAV Network Attack Scenario}

Numerous studies \cite{dai2022unmanned}\cite{8743355}\cite{gordon2019security} have outlined the importance and benefits of adapting Wi-Fi-based UAVs. UAV systems use intermediate network nodes in either a centralized or decentralized manner for data exchange. A centralized network uses a router-based approach to control and maintain network devices. On the other hand, a decentralized network uses another network node for data exchange. A centralized network node can act as a single-point vulnerability as it can support collaborative attacks such as DDoS, making it crucial to analyze \cite{unknown}. The existing works have adopted ML and DL algorithms for DDoS detection of TCP and ICMP attacks using ML and DL algorithms. However, they have not considered a mix of two DDoS flooding attacks \cite{shrivastava2023survey}.

\subsection{DDoS Attacks}

Due to increased complexity and frequency, DDoS attacks have become a prominent research problem. Lee et al. has unveiled the first taxonomy and classification details of DDoS attacks and their effects \cite{lee2004taxonomies}. These attacks can significantly compromise the network and computation resources of the target, making this calamitous for edge and UAV devices. DDoS attacks can often be characterized by fluctuating network traffic, which can be statistically analyzed \cite{borgiani2020toward}. An efficient DDoS attack usually involves network tools to generate attacks such as TCP and ICMP flooding, which ML algorithms can detect. However, these approaches can particularly fail when a hybrid DDoS attack involving a mix of TCP with ICMP flooding is considered. The current work evaluates the performance of various ML and DL algorithms for TCP, ICMP, and TCP with ICMP attack modes in a Wi-Fi-based low-cost UAV system.



\subsection{DDoS Detection}

As emphasized in several works \cite{8569300} \cite{10115435} \cite{8743355} \cite{bouhamed2021lightweight}, ML and DL have demonstrated the capabilities of efficient DDoS detection. For instance, the work by Zhang et al. \cite{8569300} has proposed DDoS detection based on the dynamic behavior of TCP and UDP protocols. The work introduces a two-step process where the first step analyzes the traffic, followed by a mathematical representation. Although efficient, the work has considered a simulation-based approach and has not been tested on a practical UAV system. The work by Cam et al. \cite{10115435} has efficiently summarised various available approaches for DDoS detection in IoT devices. The work has surveyed wavelet-based approaches followed by ML algorithms, such as Random Forest. The work has also explored DL algorithms such as autoencoders and CNN-LSTMs. However, the work has not considered the effect of longer attack sequences, which can lead to increased chances of incorrect predictions.  

The work by Tlili et al. \cite{tlili2023new} has developed a general fault detection mechanism using LSTM, Bi-LSTM, and Gated Recurrent Unit algorithms. The authors have emphasized and demonstrated the capabilities and advantages of using Bi-LSTM for multi-fault-class detection scenarios. Further, the authors have mentioned the need and necessity for studying longer sequences to detect potential anomalies in a UAV scenario. Although efficient, Vaswani et al.'s work \cite{NIPS2017_3f5ee243} has proven the unparalleled capabilities of transformers with positional encoding and multi-head self-attention mechanism can better capture the long-term dependencies than LSTM, Bi-LSTM based approaches in general. In the current work, we have adapted the Time Series Transformer (TST) \cite{10.1145/3447548.3467401}, a modified and more suitable version of the original transformer architecture for time series analysis for DDoS detection.

The rest of the paper is organized as follows. Section \ref{DDoS Attack Framework and Scenario} shows how these attacks work, ways to detect them in Section \ref{DDoS Detection}, our setup for experiments in Section \ref{Experimental Setup}, and the results we found shown in Section \ref{Experiments and Results}. Finally, in Section \ref{Conclusion and Future Scope}, we wrap up with conclusions and ideas for future research.

\section{DDoS Attack Framework and Scenario}
\label{DDoS Attack Framework and Scenario}

In the current work, three DDoS flooding attacks are considered to analyze the effect of a DDoS attack on a Wi-Fi-based UAV system. The overall system architecture and the different flooding attacks are elaborated below.

\begin{figure}[hbt]
  \centering
  \includegraphics[width=0.8\columnwidth]{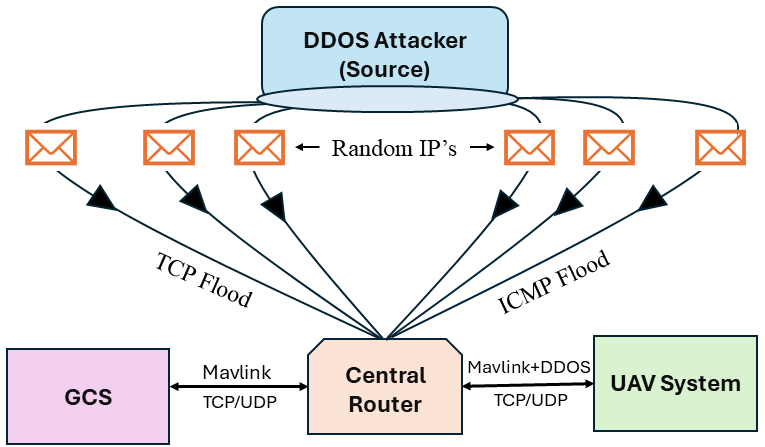}
  \caption{System Architecture for DDoS Flooding Attacks}
  \label{Figure2}
\end{figure}

\subsection{Overall Flooding Architecture}

Figure \ref{Figure2} illustrates the overall DDoS architecture with three flooding approaches: TCP flooding, ICMP flooding, and TCP with ICMP flooding. A typical DDoS attack generates numerous DoS messages through single or multiple compromised devices. Each DoS message is assigned unique Internet Protocol (IP) and Media Access Control (MAC) addresses to mask their origins. These random messages then target the UAV system's IP address, ultimately leading to compromised network bandwidth for the UAV system. The central router is the primary focal point for receiving the flood of illegitimate traffic and authentic information from the GCS. This information is then forwarded to the target UAV system.


\subsection{DDoS Flooding Approaches}
TCP flooding and ICMP flooding are common approaches used in DDoS attacks to compromise the network bandwidth. The specific and general input parameters used in the Hping3 tool (an open-source packet generator) \cite{Hping3}, for TCP and ICMP flooding are populated in Table \ref{table_flooding}.

\begin{table}[hbtp]
 \centering
 \begin{tabular}{|c|c|c|c|}
 \hline
 S.No. & Parameter & Description & Usage \\ [0.5ex] \hline
 1   & U (TCP)         & Set Urgent flag (TCP mode) & -U \\ [0.5ex] \hline
 2   & p (TCP)        & Target port & -p 80 \\ [0.5ex] \hline
 3   & 1 (ICMP)        & Set ICMP mode & -1 \\ [0.5ex] \hline
 4   & data      & Size of packet body & --data 1000 \\ [0.5ex] \hline
 5   & n         & Do not resolve hostnames & -n \\ [0.5ex] \hline
 6   & flood     & Send packets as fast as possible & --flood \\ [0.5ex] \hline
 7   & target IP & Destination IP address & 10.42.0.34 \\ [0.5ex] \hline
 \end{tabular}
 \caption{Hping3 Parameters for flooding attacks.}
 \label{table_flooding}
\end{table}

\subsubsection{TCP Flooding}
TCP flooding involves transmitting many TCP connection requests to the target system from single or multiple sources, rendering it difficult for the target system to differentiate between legitimate and malicious traffic. The description of Hping3 tool parameters used for TCP flooding attack from Table \ref{table_flooding} is as follows. Parameter $<$\texttt{U}$>$ is used to set the Urgent flag, indicating that the receiver should prioritize the data following it. Parameter $<$\texttt{p}$>$ refers to the target port address, followed by $<$\texttt{data}$>$ refers to the size of data being posted in bytes. Parameter $<$\texttt{n}$>$ ensures the host names are not resolved, increasing the attack efficiency. $<$\texttt{flood}$>$ ensures that the flooding happens with all the available system resources, and $<$\texttt{target IP}$>$ indicates the destination IP address.

\subsubsection{ICMP Flooding}
ICMP flooding involves sending a barrage of incorrectly defined ICMP packets to the target system. The target system attempts to respond to each received request, ultimately leading to exhausted network bandwidth. Leaving out the TCP-specific parameters $<$\texttt{U}$>$ and $<$\texttt{p}$>$, ICMP flooding attacks use the same parameters as TCP flooding attacks along with $<$\texttt{1}$>$ flag to indicate ICMP attack mode in the Hping3 tool.

\subsubsection{TCP With ICMP Flooding}
This hybrid attack involves transmitting many TCP connection requests to the target while simultaneously sending a barrage of incorrectly defined ICMP packets to the target system. Two parallel instances of the Hping3 tool can be used for generating this hybrid attack.

\subsection{Preliminary observations of a Flood Attack}

Table \ref{table:inter_arrival_stats} presents a 10-minute TCP and ICMP flood attack analysis. It can be observed that TCP packets have higher inter-arrival times than ICMP. This can be explained by TCP's data handshake process, which ensures data reliability with increased overhead. In contrast, ICMP is designed for speedy message exchange, which accounts for its quicker packet delivery.

\begin{table}[H]
\centering
\begin{tabular}{|l|c|c|}
\hline
\textbf{Statistic} & \textbf{ICMP} & \textbf{TCP} \\
\hline
Mean & \(5.6 \times 10^{-3}\) s & \(7.7 \times 10^{-3}\) s \\
Standard Deviation & \(4.48 \times 10^{-2}\) s & \(4.52 \times 10^{-2}\) s \\
25th Percentile & \(1.1 \times 10^{-5}\) s & \(1.2 \times 10^{-5}\) s \\
50th Percentile & \(5.55 \times 10^{-4}\) s & \(4.9 \times 10^{-5}\) s \\
75th Percentile & \(1.475 \times 10^{-3}\) s & \(4.126 \times 10^{-3}\) s \\
Arrival Rate & \(1.7857 \times 10^2\) pac/s & \(1.2987 \times 10^2\) pac/s \\
\hline
\end{tabular}
\caption{Inter-Arrival Times Statistics for Packets}
\label{table:inter_arrival_stats}
\end{table}

\section{DDoS Detection}
\label{DDoS Detection}

\begin{figure}[!hbtp]
  \centering
  \includegraphics[width=0.9\columnwidth]{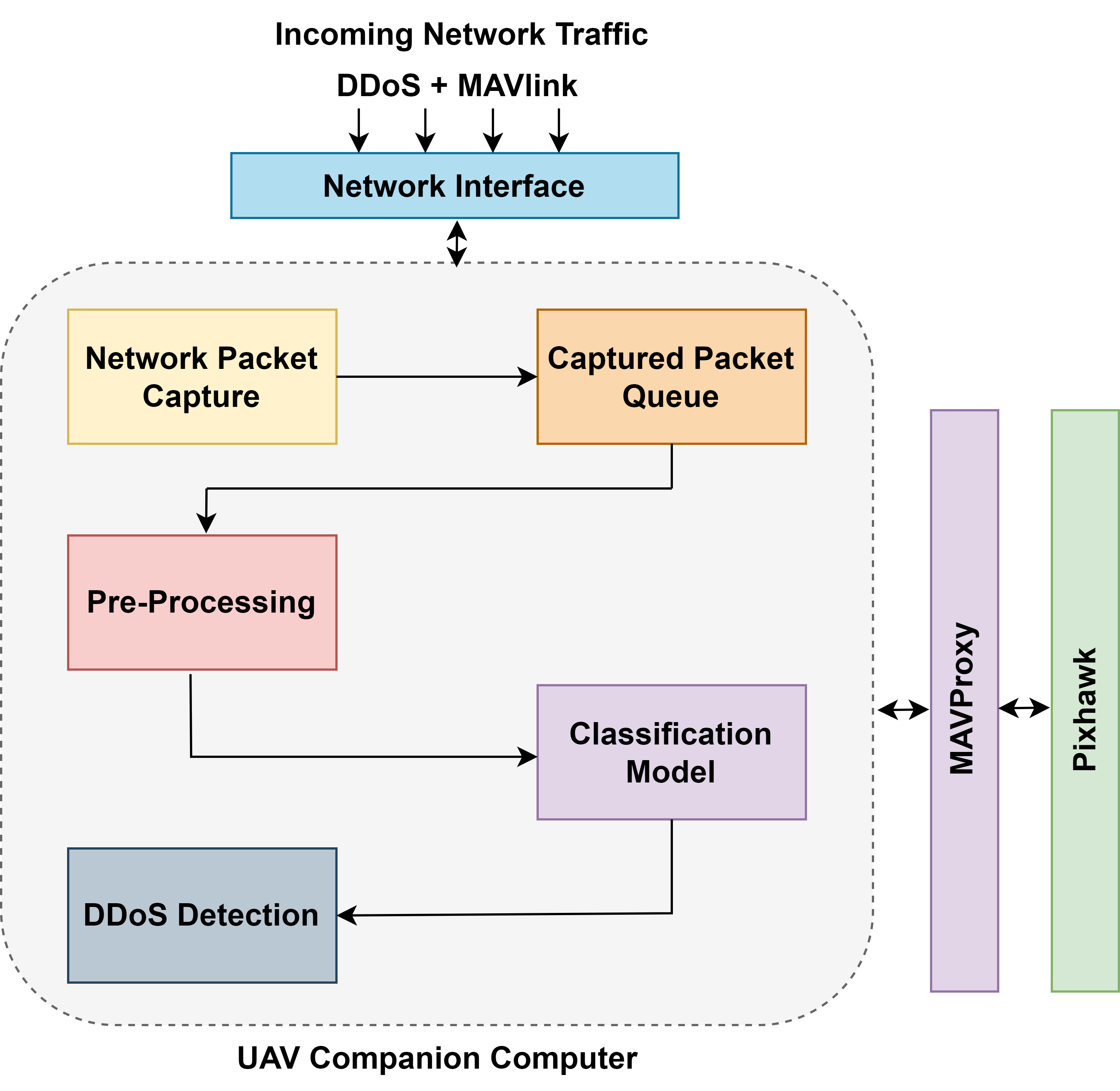}
  \caption{System Pipeline for companion computer based DDoS Detection for UAVs using Pixhawk flight controller}
  \label{Figure_DDoS_detection}
\end{figure}

In this section, we delve into the DDoS Detection mechanism that runs on the companion computer of the UAV system. Figure \ref{Figure_DDoS_detection} illustrates the pipeline for DDoS attack detection on the companion computer of the UAV system using Pixhawk flight control hardware. The network interface of the companion computer receives the MAVlink communication packets along with DDoS flood packets. All the incoming packets are then captured and forwarded to the packet queue for pre-processing, after which the classification algorithm is used for DDoS attack detection. MAVProxy, an open-source tool \cite{mavproxy}, acts as a proxy bridge to enable Wi-Fi-based data communication for the UAV system and is elaborated further in Section \ref{UAV_system_Description}

\subsection{Data Collection for Model Training and Evaluation}

The incoming network traffic is captured using Wireshark, an open-source packet analyzer \cite{WireShark}. The packet capture component of Wireshark captures the relative arrival time, packet description, protocol name, and the sequence length of the incoming traffic.

\textbf{Training Data:} The training data is collected as a continuous packet capture of normal traffic followed by a DDoS attack. Here a uniform window of 10 mins of normal network traffic data is followed by a 10 mins capture of a DDoS infested traffic. This is repeated for both the DDoS attacks separately. The total packet count and their capture duration are tabulated in Table \ref{table:train_data_description}. 

\begin{table}[ht]
\centering
\begin{tabular}{|>{\centering\arraybackslash}m{1.5cm}|>{\centering\arraybackslash}m{2cm}|>{\centering\arraybackslash}m{2cm}|>{\centering\arraybackslash}m{1.2cm}|}
\hline
\textbf{Name} & \textbf{Scenario} & \textbf{Packet Count} & \textbf{Duration (m)} \\ \hline

\multirow{4}{*}{\parbox{1.5cm}{\centering Normal + \\ TCP}} & \multirow{2}{*}{Benign} & \multirow{2}{*}{60,401} & \multirow{2}{*}{10} \\ 
                              &                      &                      &  \\ \cline{2-4} 
                              & \multirow{2}{*}{Malicious} & \multirow{2}{*}{121,401} & \multirow{2}{*}{10} \\ 
                              &                         &                      &  \\ \hline

\multirow{4}{*}{\parbox{1.5cm}{\centering Normal + \\ ICMP}} & \multirow{2}{*}{Benign} & \multirow{2}{*}{73,663} & \multirow{2}{*}{10} \\ 
                              &                      &                      &  \\ \cline{2-4} 
                              & \multirow{2}{*}{Malicious} & \multirow{2}{*}{86,929} & \multirow{2}{*}{10} \\ 
                              &                         &                      &  \\ \hline

\end{tabular}
\caption{Train Data Description}
\label{table:train_data_description}
\end{table}

\textbf{Testing Data:} Similar to train data generation, a test permutation involves 10-minute normal network traffic followed by a 10-minute DDoS traffic. Here three permutations are used to validate the performance of various DDoS detection algorithms: Normal + TCP Data, Normal + ICMP Data, and Normal + mix of TCP and ICMP Data. Table \ref{table:test_data_description} summarises these various modes along with their packet count and durations.

\begin{table}[ht]
\centering
\begin{tabular}{|>{\centering\arraybackslash}m{1.5cm}|>{\centering\arraybackslash}m{2cm}|>{\centering\arraybackslash}m{2cm}|>{\centering\arraybackslash}m{1.2cm}|}
\hline
\textbf{Name} & \textbf{Scenario} & \textbf{Packet Count} & \textbf{Duration (m)} \\ \hline

\multirow{4}{*}{\parbox{1.5cm}{\centering Normal + \\ TCP}} & \multirow{2}{*}{Benign} & \multirow{2}{*}{54,203} & \multirow{2}{*}{10} \\ 
                              &                      &                      &  \\ \cline{2-4} 
                              & \multirow{2}{*}{Malicious} & \multirow{2}{*}{110,945} & \multirow{2}{*}{10} \\ 
                              &                         &                      &  \\ \hline

\multirow{4}{*}{\parbox{1.5cm}{\centering Normal + \\ ICMP}} & \multirow{2}{*}{Benign} & \multirow{2}{*}{54,203} & \multirow{2}{*}{10} \\ 
                              &                      &                      &  \\ \cline{2-4} 
                              & \multirow{2}{*}{Malicious} & \multirow{2}{*}{165,073} & \multirow{2}{*}{10} \\ 
                              &                         &                      &  \\ \hline

\multirow{4}{*}{\parbox{1.5cm}{\centering Normal + \\ (TCP+ICMP)}} & \multirow{2}{*}{Benign} & \multirow{2}{*}{36,025} & \multirow{2}{*}{10} \\ 
                              &                      &                      &  \\ \cline{2-4} 
                              & \multirow{2}{*}{Malicious} & \multirow{2}{*}{441,442} & \multirow{2}{*}{10} \\ 
                              &                         &                      &  \\ \hline                              

\end{tabular}
\caption{Test Data Description}
\label{table:test_data_description}
\end{table}


\subsection{Data Pre-Processing}

\begin{figure}[!hbtp]
  \centering
  \includegraphics[width=1.0\columnwidth]{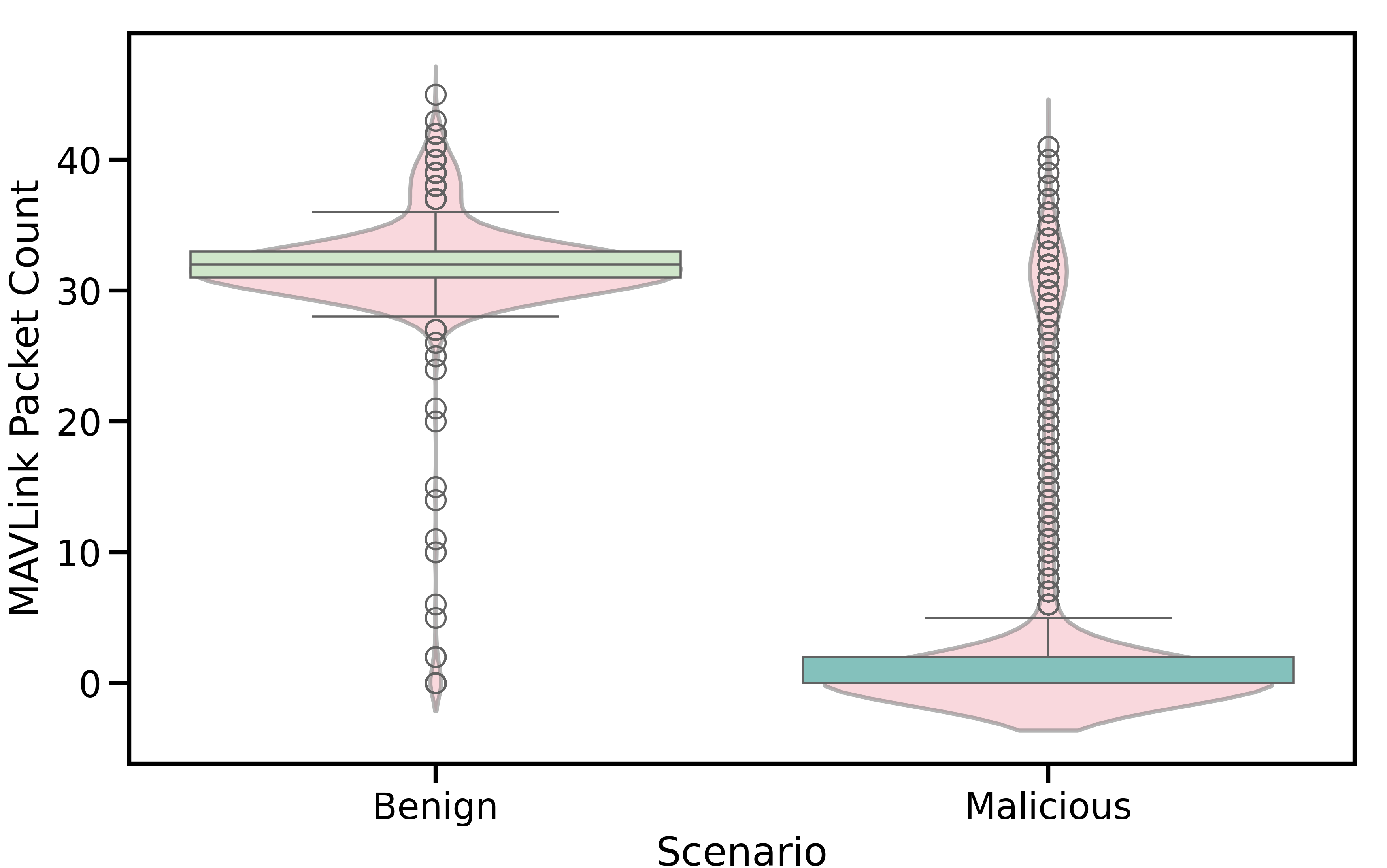}
  \caption{MAVLink Packet count in a DDoS attack}
  \label{Figure_tcp_box}
\end{figure}

In the current work, the MAVLink packet count is used as the classification feature to identify potential DDoS attacks. A sliding window of 0.1 seconds is considered to aggregate the number of MAVLink packets arrived based on relative arrival times. The aggregated data is then binary encoded with zero as the normal condition and one as the DDoS condition. Figure \ref{Figure_tcp_box} illustrates the variation in total MAVLink packet count over a window of 0.1 seconds. It can be observed that MAVLink packets decrease significantly during a DDoS attack, thereby making it a resourceful indicator for classification. However, a time series sequence in the variation of MAVlink packet count can better identify a malicious scenario.

Keeping the training data, which has normal, TCP, and ICMP traffic information as a reference, Standard Scaler is applied to each of the testing conditions, where the mean value is subtracted from each of the test conditions, and then the result is divided by the standard deviation of the data.
\subsection{Proposed and Comparision Algorithms to Detect DDoS}

Keeping, the data pre-processing approach the same, six algorithms: XGBoost, Isolation Forest, LSTM, Bi-LSTM, LSTM with attention, and Bi-LSTM with attention are adopted as the comparison algorithms to evaluate the classification performance of our proposed Time Series Transformer. The description and the necessary hyper-parameters of these models are elaborated further.

\textbf{XGBoost:} Extreme Gradient Boosting is a supervised learning algorithm that uses a gradient boosting approach to minimize the overall training loss. XGBoost is a tree-based linear model optimized for classification problems \cite{xgboost}. Grid Search is used for hyper-parameter tuning and tuned parameters are: \textit{n\_estimators} equal to 100, \textit{max\_depth} equal to 4, and \textit{subsample} equal to 1. \textit{n\_estimators} indicates the number of tree estimators for classification, \textit{max\_depth} indicates the tree depth and  \textit{subsample} indicates that each estimator is trained on whole dataset.





\textbf{LSTM and Bi-LSTM Attention Models:} In this study, we use LSTM and its extensions, including Bi-LSTM equipped with an attention mechanism to process Mavlink count data. LSTMs excel in retaining information from the Mavlink count data over extended periods, making them adept at modeling sequential information. Bi-LSTMs expand upon this by processing the input sequence in both forward and reverse order, thereby enriching the model's understanding with additional context, which can significantly boost performance in sequence classification tasks \cite{10136944}. Integrating an attention mechanism \cite{attn} further refines the model's capability by focusing on crucial features within the Mavlink count data, enabling the selective emphasis of important information and highlighting its superior ability to detect DDoS attacks. Our attention mechanism highlights crucial information by adjusting weights, taking the dot product of weights and inputs, adding bias, and then applying a tanh function followed by a softmax layer to refine the focus on significant features. In this work, we focused on optimizing the models by tuning the sequence length, the number of layers, the number of units in the dense layer, and the dropout rate.

\subsection{Time Series Transformer:}
\begin{figure}[hbt]
 \centering
 \includegraphics[width=0.60\columnwidth]{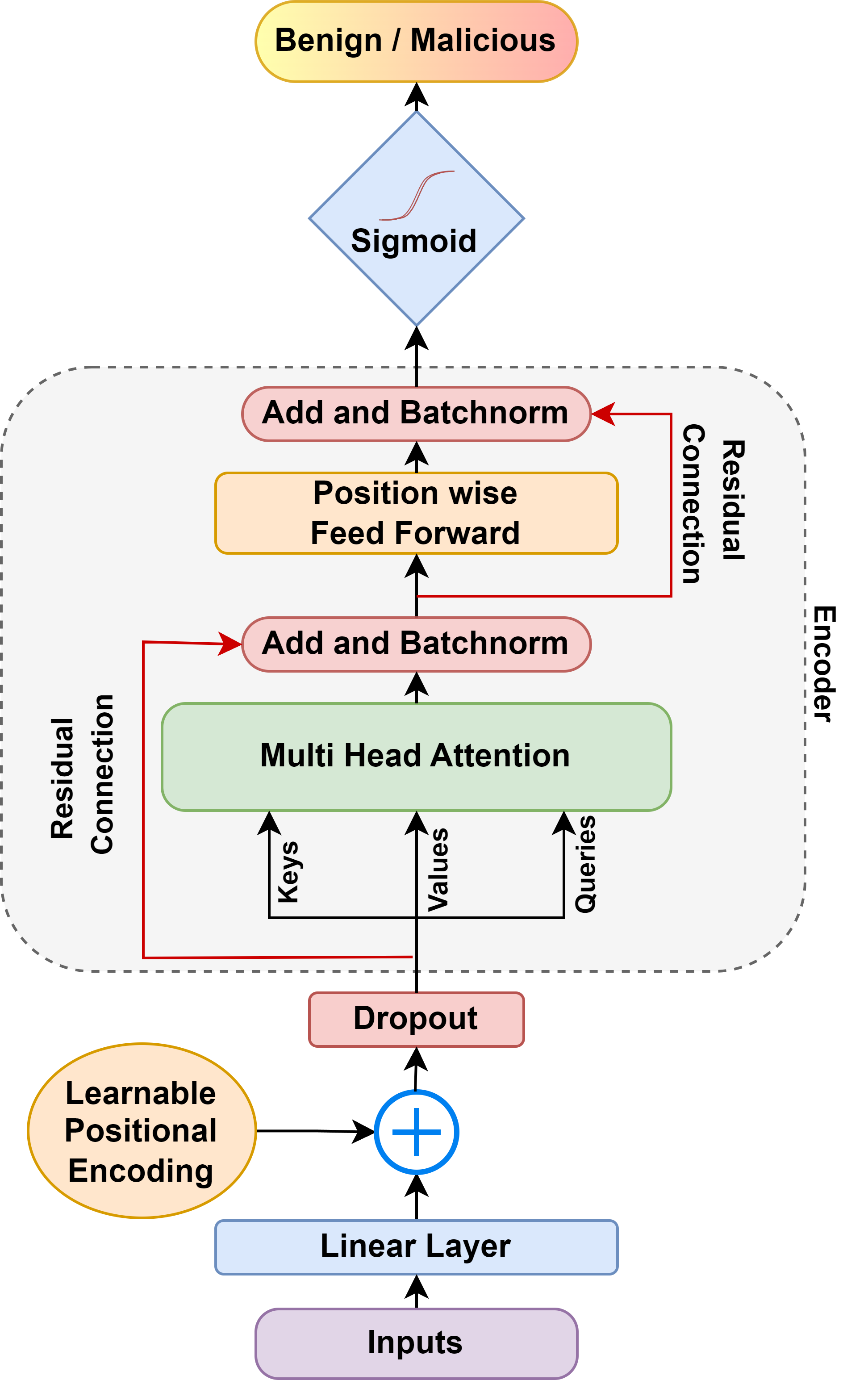}  \caption{Architecture of Time Series Transformer}
 \label{TST Architecture}
  \vspace{-0.3cm}
\end{figure}

The TST by George et al. \cite{10.1145/3447548.3467401} has generalized transformer architecture for time series analysis. Unlike the baseline transformer architecture by Vaswani et al. \cite{NIPS2017_3f5ee243}, this uses an encoder-only approach to perform classification and regression analysis. Figure \ref{TST Architecture} illustrates the currently adapted TST architecture. The input data sequence is projected by a 1D convolution layer to match the dimension of the encoder layer (\textit{d\_model}). In contrast to the sinusoidal embeddings in the original transformer, here, fully learnable positional encodings are added to the output of the 1D convolution layer, which is given as the keys, queries, and values to the multi-head self-attention layer. Multi-head self-attention helps retain the longer sequences of DDoS data. The output of the attention layer is batch-normalized, after which it is provided as an input to the feed-forward network. A sigmoid function is used to threshold the output for classification.

In the current work, We have considered six parameters namely: \textit{seq\_len}, \textit{d\_model}, \textit{n\_heads}, \textit{d\_ff}, \textit{n\_layers}, and \textit{dropout} for tuning the classification performance of the TST model. The parameter \textit{seq\_len} determines the length of the input sequence given as an input. Since each entry in the data is created at 0.1 seconds, a sequence length of 400 would translate to past 40 seconds of variations in MAVLink packet count. The parameter \textit{d\_model} determines the dimension of the input data to the encoder model, \textit{n\_heads} is the number of attention head elements given as input to create the attention vector, \textit{d\_ff} is the size of the feed-forward network which is present before the classification head, \textit{n\_layers} is the number of cascaded encoder layers before the classification head, and \textit{dropout} is used to randomly prune the transformer to prevent over-fitting during the training phase. The rest of the hyper-parameters are listed in Table \ref{table:tst}. Section \ref{Ablation} elaborates on the parameter ablations and analysis.

\begin{table}[ht]
\centering
\begin{tabular}{|>{\centering\arraybackslash}m{0.8cm}|>{\centering\arraybackslash}m{1.2cm}|>{\centering\arraybackslash}m{2.5cm}|>{\centering\arraybackslash}m{1.2cm}|}
\hline
\textbf{S.No.} & \textbf{Parameter} & \textbf{Description} & \textbf{Parameter Value} \\ \hline

1 & seq\_len & Length of I/P sequence & 400 \\ \hline

2 & d\_model & Dimension of Encoder & 64 \\ \hline

3 & n\_heads & Number of Attention Heads & 32 \\ \hline

4 & d\_ff & Dimension of Feed-Forward N/W & 64 \\ \hline

5 & n\_layers & No. of Encoder Layers & 2 \\ \hline

6 & dropout & Dropout in Encoder and FF layer & 0.1 \\ \hline

7 & bs & Batch Size & 4 \\ \hline

8 & d\_k & Key Dimensions & 2 \\ \hline

9 & d\_v & Values Dimensions & 2 \\ \hline

10 & le & Learned Embeddings & True \\ \hline

11 & act & activation & gelu \\ \hline

12 & seed & random seed value & 42 \\ \hline

\end{tabular}
\caption{Hyperparameter of TST Algorithm}
\label{table:tst}
\end{table}

\section{Experimental Setup}
\label{Experimental Setup}

This section sheds light on various hardware and software components used in the current work for DDoS Attack and Detection. The description of the UAV system, Hardware, and Software components are as follows. 

 \begin{figure}[hbtp]
    \centering
    \begin{subfigure}[c]{0.7\columnwidth}
        \includegraphics[scale=0.13]{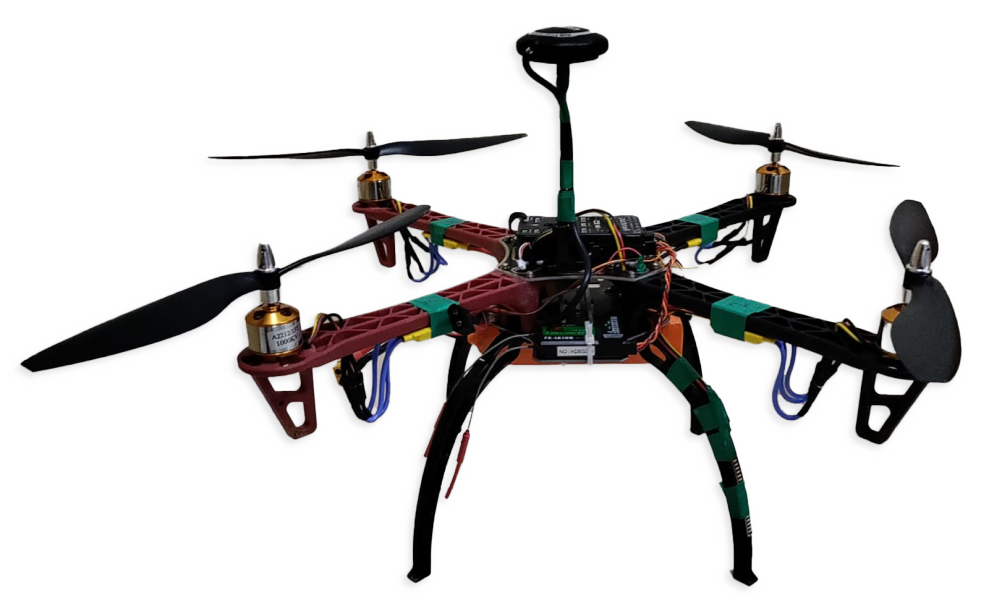}
        \caption{Custom Built UAV System}
	   \label{Figure:UAV}
    \end{subfigure}
    
    
 	\begin{subfigure}[t]{.23\textwidth}
		\includegraphics[scale=0.070]{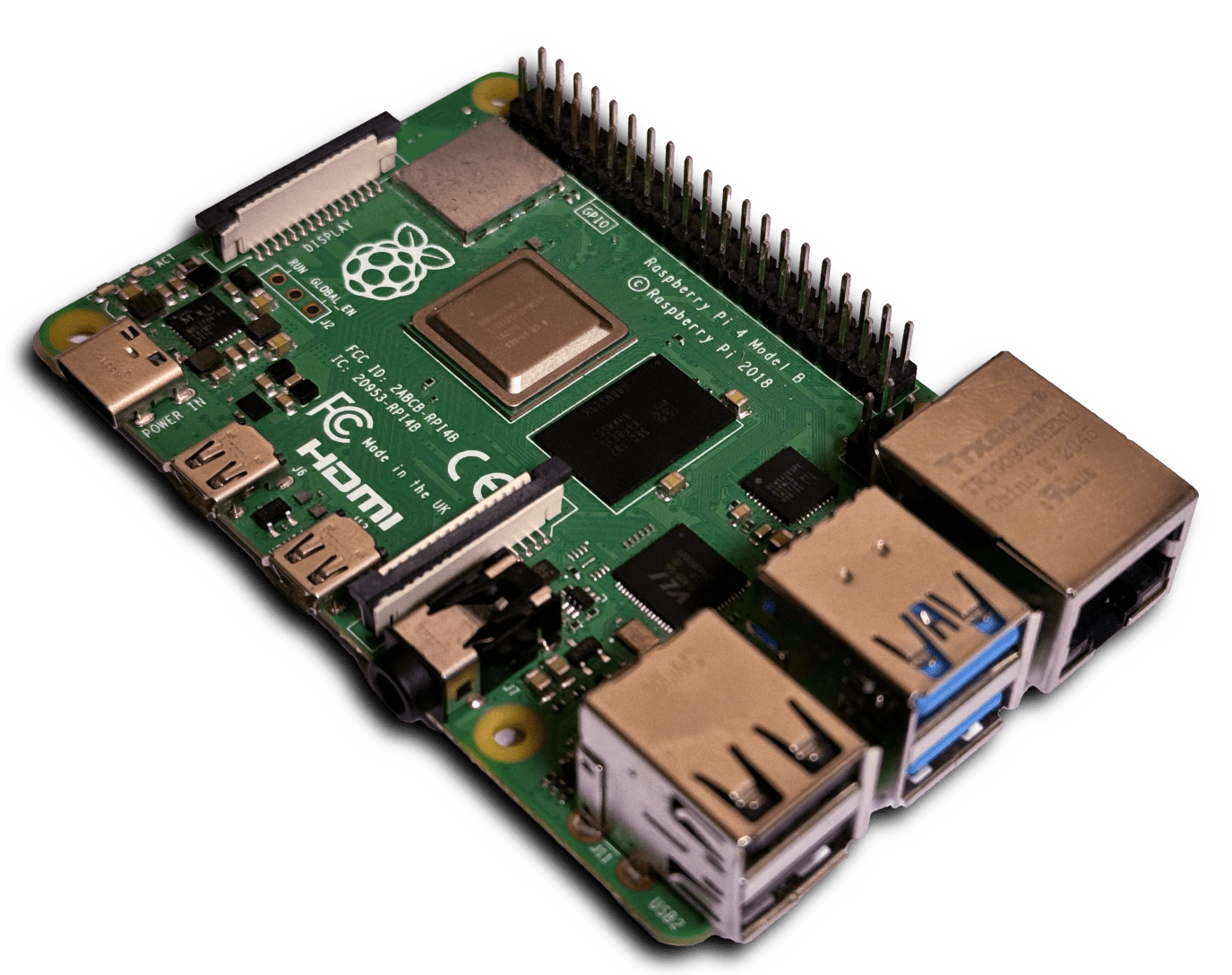} 
		\caption{Companion Computer}
		\label{fig:RPI}
	\end{subfigure}
     \begin{subfigure}[t]{.23\textwidth}
		\includegraphics[scale=0.070]{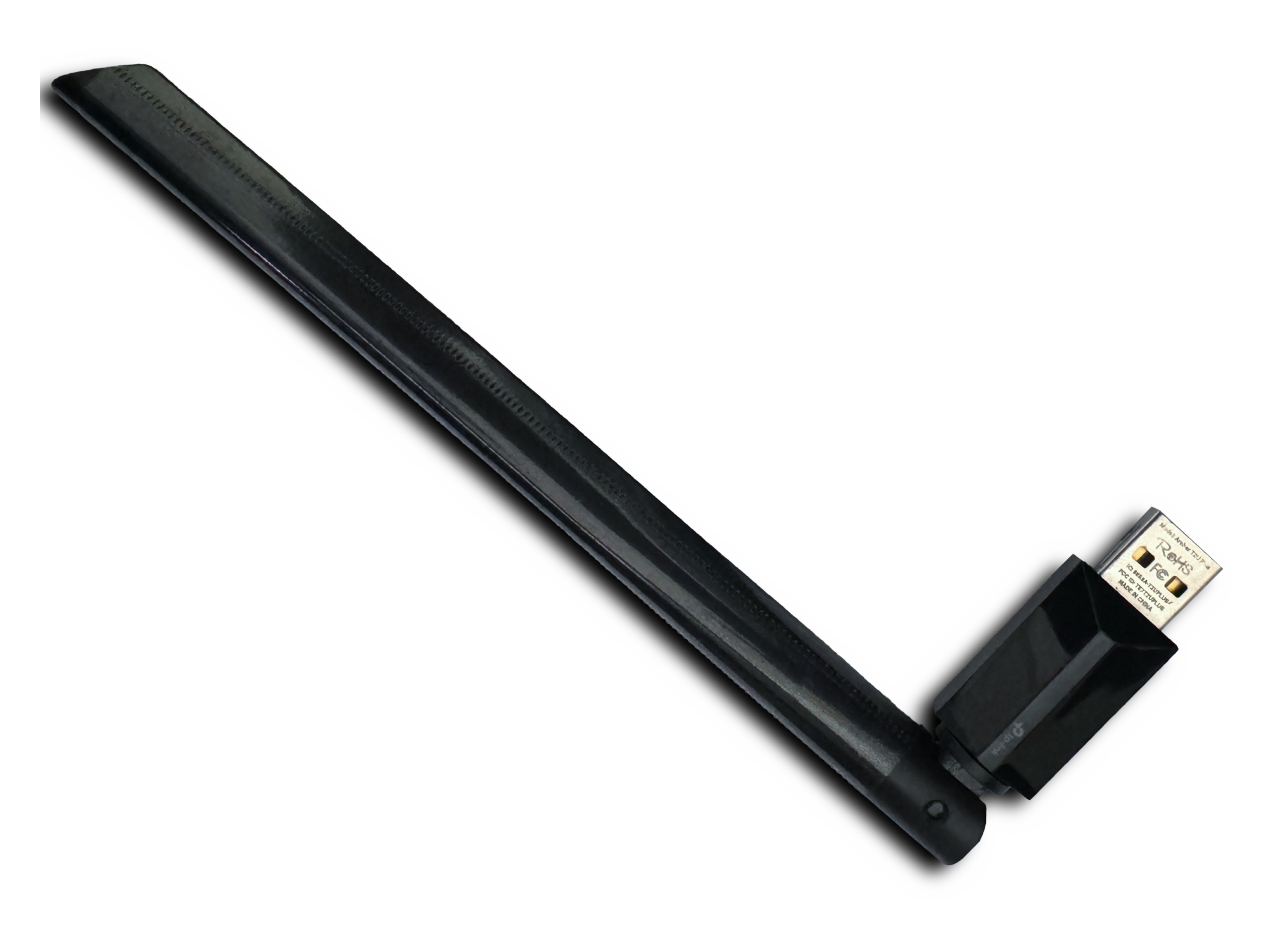} 
		\caption{External Wi-FI Adapter}
		\label{fig:Antenna}
	\end{subfigure}

  \caption{Hardware Components in the Experimental Setup}
	\label{Figure_experimental}
\end{figure}

\subsection{UAV System Description}
\label{UAV_system_Description}

Figure \ref{Figure:UAV} illustrates the different onboard components of the custom-built quad-copter used in the current work. PixHawk 2.4.8 is used as the flight controller. It provides various sensors capable of sensing positional information through the integrated IMU and magnetometer. The other onboard components include a Radio Telemetry for control information exchange with the GCS and a UBlox Neo-M8N Module \cite{10260534}. 

The UAV System includes Raspberry Pi 4B \cite{rpi4b}  as an onboard companion computer. MAVProxy, an open-source GCS, acts as a proxy bridge to enable Wi-Fi-based data communication for the UAV system and also enables companion computing \cite{mavproxy}. MAVProxy uses MAVlink 2.0 protocol encapsulated in TCP/UDP Header for data communication between UAV system and GCS.  


\subsection{Hardware Components}
The various hardware components used in the current work are as follows

\textbf{GCS and Attacker Systems:} GCS and Attacker systems use identical configurations. The system is a Ryzen 7 5825U Octa-core processor with Radeon Graphics clocked at 2000 MHz. The system runs with 16 GB of RAM and 512 GB of Solid-State Storage, and an 802.11ac WiFi Transceiver.

\textbf{Companion Computer:} As illustrated in Figure \ref{fig:RPI}, Raspberry Pi 4B, a general purpose and versatile system, is chosen as the Companion Computer and the Central Router. This 64-bit Quadcore SoC runs at 1.5GHz with support for 802.11 b/g/n/ac standards \cite{rpi4b} \cite{gudla2018defense}. 

\textbf{External Wi-Fi Adapter:} Figure \ref{fig:Antenna} displays a TP-link AC600 high gain wireless dual band usb adapter \cite{tplink}. This adapter facilitates a peak transfer rate of 200 Mbps with a 5dBi high-gain antenna, enabling powerful DDoS traffic injection.

\subsection{Software Componenets}
The various software components used in the current work are as follows

\textbf{Wireshark: } Wireshark is an open-source network packet analyzer. This tool enables users to monitor network traffic in real-time. This tool is mainly used for DDoS data generation for the model training phase in the current work \cite{WireShark}. 

\textbf{HPing3: } Hping3 is an open-source packet generator tool capable of producing a network flooding attack. This tool aims to generate sufficient network packets leading to compromised network bandwidth \cite{Hping3}. 

\textbf{QGroundControl: } Q Ground-Control, a user-friendly
GCS is adapted for analyzing the UAV state and for control \cite{10260534}.

\section{Experiments and Results}
\label{Experiments and Results}
In this section, we evaluate the classification performance of various algorithms for DDoS detection. We then present the memory utilization and inference time for these algorithms. Following that, we present ablation study results to evaluate the best hyperparameters for the TST. 

All the algorithms are trained and tested on the Colab notebook using Nvidia T4 GPU. Inference time for the algorithms is calculated on RPI-4B 8 GB RAM. The various results are as follows

\subsection{Performance Evaluation}

In this section, we present the performance evaluation of
XGB, IF, LSTM, Bi-LSTM, LSTM-A, BLSTM-A for three
kinds of flooding attacks TCP, ICMP and TCP with ICMP
as shown in Table II. The model’s effectiveness is evaluated
using performance metrics including F1 score, accuracy,
recall, and precision.

\begin{table}[htbp]
\setlength\tabcolsep{-1pt}
\centering
\begin{tabular*}{0.9\linewidth}{@{\extracolsep{\fill}} l cccc }
            \hline 
                   &  & TCP Flooding & &  \\ \hline

            & Precision & Recall & Accuracy & F1 score \\ \hline
            XGB  &0.90      &0.96    &0.94  &0.92 \\ 
            IF  &0.68      &0.76  &0.67    &0.64 \\  
            LSTM  &0.93  &0.97   &0.96   &0.95 \\  
            Bi-LSTM  &0.94  &0.96   &0.97   &0.95 \\ 
            LSTM-A  &0.94  &0.96   &0.97   &0.95 \\  
            BLSTM-A  &0.99  &0.98   &0.99   &0.99 \\ 
            TST &\textbf{0.99}  &\textbf{0.99}   &\textbf{0.99}   &\textbf{0.99} \\ 
            \hline
           
    \end{tabular*}
    \caption{Performance evaluation for TCP Flooding}
    \label{table:tcp_table}
\end{table}

\textbf{TCP Flooding:} Table \ref{table:tcp_table} presents the performance for TCP Flooding attack. The significant deviation in classification accuracy of XGB, IF, and other algorithms can be explained by the time series nature of LSTMs, and TST-based algorithms. It can be observed that the adoption of the attention mechanism has significantly improved TCP Flood detection owing to the extended capabilities of the attention mechanism to analyze longer sequences. TST algorithm stands out with maximum performance across all the metrics due to its capability to handle longer sequences with learned positional embeddings.

\begin{table}[htbp]
\setlength\tabcolsep{-1pt}
\centering
\begin{tabular*}{0.9\linewidth}{@{\extracolsep{\fill}} l cccc }
            \hline 
                   &  & ICMP Flooding & &  \\ \hline

            & Precision & Recall & Accuracy & F1 score \\ \hline
            XGB  & 0.89      &0.95   &0.93   &0.91 \\ 
            IF  &0.65      &0.72  &0.61    &0.60 \\  
            LSTM  &0.71  &0.92   &0.85   &0.75 \\  
            Bi-LSTM  &0.76  &0.95   &0.90   &0.81 \\ 
            LSTM-A  &0.78  &0.95   &0.91  &0.84 \\  
            BLSTM-A  &0.89  &0.98   &0.97   &0.93 \\ 
            TST &\textbf{0.99}  &\textbf{0.99}   &\textbf{0.99}   &\textbf{0.99} \\ 
            \hline
           
    \end{tabular*}
    \caption{Performance evaluation for ICMP Flooding}
    \label{table:icmp_table}
\end{table}

\textbf{ICMP Flooding:} Table \ref{table:icmp_table} presents the performance for ICMP Flooding attack. A similar trend, such as TCP flooding evaluation, can be observed for ICMP flood detection. However, unlike TCP flooding, the Bi-LSTM-A algorithm drastically decreases classification efficiency, while TST presents a consistent performance. This can be explained by multiple encoder layers of the TST architecture adopted in the current work. Multiple encoder layers can efficiently help identify the hidden trends in the data. 

\begin{table}[htbp]
\setlength\tabcolsep{-1pt}
\centering
\begin{tabular*}{0.9\linewidth}{@{\extracolsep{\fill}} l cccc }
            \hline 
                   &  & TCP + ICMP Flooding & &  \\ \hline

            & Precision & Recall & Accuracy & F1 score \\ \hline
            XGB  &0.74      &0.90   &0.83   &0.77 \\ 
            IF  &0.67  &0.81   &0.70   &0.65 \\  
            LSTM  &0.69  &0.90   &0.82   &0.72 \\  
            Bi-LSTM  &0.72  &0.92   &0.86   &0.76 \\ 
            LSTM-A  &0.74  &0.94   &0.88   &0.79 \\  
            BLSTM-A  &0.79 &0.96   &0.92   &0.85 \\ 
            TST &\textbf{0.92}  &\textbf{0.98}   &\textbf{0.97}   &\textbf{0.94} \\ 
            \hline
           
    \end{tabular*}
    \caption{Performance evaluation for TCP+ICMP Flooding}
    \label{table:tcp_icmp_table}
\end{table}

\textbf{TCP + ICMP Flooding:} TST algorithm follows a consistent efficiency in classifying a hybrid DDoS scenario involving TCP and ICMP attacks as shown in Table \ref{table:tcp_icmp_table}. This unparalleled performance can be explained by the capabilities of transformer architecture to better capture the trends and variations in the data due to the presence of the multi-head self-attention along with the multiple encoders, which are absent in other algorithms.

\subsection{Analysis of Memory and Prediction Time}

\begin{table}[!hbtp]
\centering
\begin{tabular}{|>{\centering\arraybackslash}m{0.8cm}|>{\centering\arraybackslash}m{1.4cm}|>{\centering\arraybackslash}m{2.5cm}|>{\centering\arraybackslash}m{1.2cm}|}
\hline
\textbf{S.No.} & \textbf{Algorithm} & \textbf{Memory(MB)} & \textbf{Inference Time (s)} \\ \hline

1 & XGB & 7.14 & 0.0005  \\ \hline

2 & IF & 13.89 & 0.0001  \\ \hline

3 & LSTM & 253.62 & 0.0005 \\ \hline

4 & Bi-LSTM & 255.71 & 0.0022 \\ \hline

5 & LSTM-A & 257.54 & 0.0019 \\ \hline

6 & BLSTM-A & 259.94 & 0.0061 \\ \hline

7 & TST & 371.32 & 0.1339 \\ \hline

\end{tabular}
\caption{Computational analysis of test data on RPI-4B}
\label{table:mem&inf}
\end{table}

We analyzed the memory usage and inference time for each algorithm on the companion computer of the UAV system. Table \ref{table:mem&inf}  shows that algorithms with attention mechanisms require more memory and inference time than their counterparts without attention, as the mechanism requires additional computation to handle the input data. While the TST model outperforms the other algorithms, it incurs a higher memory and inference time due to the presence of multiple encoder layers for DDoS detection. Although TST has the highest inference time, the practical implication can be sustainable in a flight scenario as the inference time is nearly 0.1 seconds.

\subsection{TST Ablation Study}
\label{Ablation}

\begin{table}[htb]
\centering
\begin{tabular}{|c | c | c | c | c | c |}
\hline

{\multirow{2}{*}{Sequence Length}} & \multicolumn{2}{|c|}{F1 score}  & {\multirow{2}{*}{Encoder Size}} & \multicolumn{2}{|c|}{F1 score} \\
\cline{2-3} \cline{5-6}
 & TCP & ICMP &  & TCP & ICMP  \\
\hline
      100 & 0.96 & 0.89  & \textbf{64} & \textbf{0.99} & \textbf{0.99}  \\
       200 & 0.80 & 0.85   & 128 & 0.81 & 0.84 \\
       300 & 0.90 & 0.91    & 256 & 0.84 & 0.71 \\
       \textbf{400} & \textbf{0.99} & \textbf{0.99}   & 512 & 0.99 & 0.98 \\
       500 & 0.86 & 0.83   &  &  &  \\
\hline
\end{tabular}
\caption{F1 score vs Sequence Length, F1 score vs Encoder Size} 
\label{total_cell1}
\end{table}

\vspace{-0.5cm}

\begin{table}[htb]
\centering
\begin{tabular}{|c | c | c | c | c | c | c | c | c |}
\hline

{\multirow{2}{*}{Num of heads}} & \multicolumn{2}{|c|}{F1 score}  & {\multirow{2}{*}{FF size}} & \multicolumn{2}{|c|}{F1 score} \\
\cline{2-3} \cline{5-6}
 & TCP & ICMP &  & TCP & ICMP \\
\hline
      8 & 0.99 & 0.98  & 32 & 0.97 & 0.96   \\
      16 & 0.98 & 0.97 & \textbf{64} & \textbf{0.99} & \textbf{0.99}\\
      \textbf{32} &\textbf{ 0.99} & \textbf{0.99} & 128 & 0.97 & 0.98 \\
\hline
\end{tabular}
\caption{F1 score vs Num of heads, F1 score vs Feed forward size} 
\label{total_cell2}
\end{table}

\vspace{-0.5cm}

\begin{table}[htb]
\centering
\begin{tabular}{|c | c | c | c | c | c | c | c | c |}
\hline

{\multirow{2}{*}{Num of encoders}} & \multicolumn{2}{|c|}{F1 score}  & {\multirow{2}{*}{Dropout}} & \multicolumn{2}{|c|}{F1 score} \\
\cline{2-3} \cline{5-6}
 & TCP & ICMP &  & TCP & ICMP \\
\hline
     1 & 0.96 & 0.96  & \textbf{0.1} & \textbf{0.99} & \textbf{0.99}     \\
    \textbf{2} & \textbf{0.99} & \textbf{0.99}  & 0.2 & 0.98 & 0.97 \\
      3 & 0.96 & 0.90   & 0.3 & 0.97 & 0.97 \\
\hline
\end{tabular}
\caption{F1 score vs Num of encoders, F1 score vs Dropout}
\label{total_cell3}
\end{table}

Table \ref{total_cell1}, \ref{total_cell2}, and  \ref{total_cell3} present the variation in sequence length and encoder size. TST model goes through a transitional phase from a sub-optimal
F1 score to an optimal score at 400 sequence length, which can be explained by the dependency of DDoS detection on longer sequences. TST reaches the optimal F1 score for a smaller encoder and feed-forward network size of 64, which the simplicity of input data can explain. The need for attention mechanisms in longer sequences for DDoS detection can explain the higher number of attention heads and encoder layers. The dropout rate reaches the optimal F1 score at a lower value of 0,1 owing to a smaller feed-forward network and the encoder size. Table \ref{table:tst} presents the optimal hyper-parameters of the proposed TST algorithm for efficient DDoS detection.


\subsection{Effect of Learnable Positional Embedding}
\label{PSE}

Table \ref{table:tst_lpe} highlights the advantage of adapting learnable positional encoding (LPE) in TST for a DDoS detection scenario. LPE enables the TST algorithm to better capture the sequential relationships between input data elements using learned weights for each sequence component. LPE can help better adjust the attention weights, ultimately determining the classification outcome.

\begin{table}[htbp]
\setlength\tabcolsep{-1pt}
\centering
\begin{tabular*}{0.9\linewidth}{@{\extracolsep{\fill}} l cccc }
            \hline 
                   &  &Effect of Learnable & &  \\ 
                   
                   &  &Positional Embeddings & &  \\ \hline
            & Precision & Recall & Accuracy & F1 score \\ \hline
            with LPE  &0.99  &0.99   &0.99   &0.99 \\ 
            without LPE &0.91  &0.95   &0.96   &0.94 \\ 
            \hline
           
    \end{tabular*}
    \caption{Effect of Learnable Positional Embeddings}
    \label{table:tst_lpe}
\end{table}

\section{Conclusion and Future Scope}
Low-cost UAV systems have become prominent due to their ease of use. A typical low-cost UAV uses Wi-Fi-based information exchange, making them vulnerable to various cybersecurity attacks such as Distribute Denial of Service (DDoS) attacks. In the current work, we first present the suitable steps and pre-processing to handle DDoS network data. We have then evaluated various ML and DL algorithms such as XGBoost, Isolation Forest, Long Short-Term Memory (LSTM), Bidirectional-LSTM (Bi-LSTM), LSTM with attention, Bi-LSTM with attention, and Time Series Transformer (TST) for DDoS detection considering three variants of DDoS attacks namely: TCP, ICMP, and TCP + ICMP attacks. Our evaluation indicates that the proposed TST model outperforms the other ML and DL algorithms. TST has demonstrated an F1 score of 0.999, 0.997, and 0.943 for TCP, ICMP, and TCP + ICMP flooding attacks. We have also conducted a TST ablation analysis for fine-tuning the hyperparameters and we have also emphasized the advantage of adapting learnable positional embeddings in TST for DDoS detection. We plan to extend this work to build a DDoS mitigation algorithm for various UAV flight scenarios as part of our future work. 


\label{Conclusion and Future Scope}


\bibliographystyle{ieeetr}
\bibliography{main}

\end{document}